\documentclass[%
 reprint,
 floatfix,
 amsmath,amssymb,
 aps,
]{revtex4-1}

\usepackage{graphicx}
\usepackage{dcolumn}
\usepackage{bm}

\begin{document}
\preprint{APS/123-QED}

\title{Giant fluctuations in the flow of fluidised soft glassy materials: an elasto-plastic modelling approach}

\author{Magali Le Goff}
\email{magali.le-goff@univ-grenoble-alpes.fr}
\address{%
Universit\'e Grenoble Alpes, CNRS, LIPhy, F-38000 Grenoble, France
}
\author{Eric Bertin}
\address{%
Universit\'e Grenoble Alpes, CNRS, LIPhy, F-38000 Grenoble, France
}
\author{Kirsten Martens}%
\address{%
Universit\'e Grenoble Alpes, CNRS, LIPhy, F-38000 Grenoble, France
}

\date{\today}

\begin{abstract}
In this work we study the rheological features of yield stress materials that exhibit non-homogeneous steady flows and that are subjected to an additional mechanical noise.
Using a mesoscale elasto-plastic model accounting for a viscosity bifurcation in the flow response to an external shear stress, we find that additional sources of noise can lead to a fluidisation effect.
As we increase the noise intensity we evidence a transition from a non-monotonic to a monotonic rheology, associated with giant fluctuations of the macroscopic shear rate and long-time correlated dynamics.
Although distinct noise models can lead to different rheological behaviours in the low stress regime, we show that the noise-induced transition from shear-localised to homogeneous flow at higher stresses appears very generic.
The observed features in the dynamics can be interpreted as a result of an out-of-equilibrium phase transition, for which we estimate the critical exponents that appear to be independent of the specific choice of the noise implementation for the microscopic dynamics.
\end{abstract}

\maketitle

\section{Introduction}
Dense disordered materials such as emulsions, foams, colloidal suspensions, or granular materials exhibit rich rheological behaviours. 
These materials have in common that once deformed beyond their solid elastic regime they yield plastically towards a complex flow regime with a shear rate dependent viscosity \cite{bonn2017yield, nicolas2018deformation}.
The steady state flow behaviour of these yield stress fluids (YSF) can be described at the continuum level using empirical laws such as the Herschel-Bulkley relationship \cite{herschel1926konsistenzmessungen}, or  continuum descriptions, such as visco-elasto-plastic \cite{marmottant2007elastic, saramito2007new} and fluidity models \cite{bocquet2009kinetic, fielding2014shear}.
While such descriptions account well for the average flow behaviour at a coarse grained scale, it has appeared that some flow features of YSF are dominated by giant fluctuations of the relevant rheological quantities on the macroscopic scale \cite{coussot2002avalanche, lootens2003giant, cantat2006stokes, bares2017local}. This can for example lead to non-local, strongly system-size dependent, transport coefficients for the material dynamics \cite{lemaitre2009rate, martens2011connecting, tyukodi2018diffusion}. Accordingly, understanding the role of mechanical noise and its spatio-temporal features has not only attracted a strong fundamental interest \cite{nicolas2018deformation} but is also of direct importance in rheological applications \cite{bonn2017yield}.

Part of the mechanical fluctuations in driven disordered materials are usually generated by the flow itself, for example resulting from the elastic response of the material to localised plastic events \cite{argon1979plastic, schall2007structural}. They are therefore very different in nature from thermally generated fluctuations \cite{nicolas2014rheology} and must be incorporated into modelling approaches in a self-consistent manner \cite{hebraud1998mode, agoritsas2015relevance}.
%
Interestingly, flow-induced fluctuations can be associated in some cases with a self-fluidisation of the material, i.e.~a decrease in shear stress with increasing shear rate. This leads to non-monotonic rheological constitutive curves \cite{schall2010shear, mansard2011kinetic, fielding2014shear}, that can be associated with flow instabilities, potentially leading to shear localisation, metastability and hysteresis \cite{wortel2014rheology}.
In the case of granular materials, this self-fluidisation process finds its origin in sliding frictional contacts \cite{wortel2014rheology,wortel2016criticality,degiuli2017friction}. In non frictional YSFs, non monotonic flow curves can be explained by mechanisms such as inertia \cite{nicolas2016effects,karimi2016role, vasisht2018permanent} or local softening following structural rearrangements \cite{coussot2010physical, martens2012spontaneous}. 
%

Besides this self-generated mechanical noise there can be additional external sources of noise, which can be regarded as a first approximation as independent of the shear-induced one. This is for example the case of thermally induced \cite{chattoraj2010universal,ikeda2012unified} or strong vibration activated irreversible deformations \cite{d2003observing,Caballero-Robledo2009,jia2011elastic,hanotin2012vibration}. Other mostly rate-independent fluctuations can also result from local processes such as coarsening in foams \cite{cohen2004origin}, or an internal activity \cite{mandal2016active,tjhung2017discontinuous,matoz2017nonlinear}.
One important aspect of such external noise sources is that they can induce a fluidisation of the system at small imposed external stresses.

An even more interesting case is the scenario where fluctuations result from the interplay of a self-fluidising and an external source of noise.
Upon an increase of the external noise magnitude, it will dominate at some critical value over the self-fluidising noise. This scenario can induce a change from a non-monotonic constitutive behaviour to a simple monotonic one, due to the fluidisation effect of the external noise. The associated transition has been investigated in the context of non-equilibrium phase transitions, both experimentally in the case of frictional granular materials \cite{wortel2016criticality} and theoretically in a generic elasto-plastic model \cite{legoff2019critical}. 
%

In this work, we consider two different models for an external fluidising noise, and show that although they lead to distinct rheological behaviours at low stress, both induce a transition from a non monotonic to a monotonic flow curve, associated with the transition from shear banded to homogeneous flow.
We evidence that the competition between the endogenous noise and an external fluidising noise leads to giant fluctuations in the flow of soft glassy materials, that become relevant on the rheological scale.
When interpreting the transition between the self-fluidised and the externally fluidised regimes upon increasing noise magnitude as a critical phenomenon \cite{wortel2016criticality, legoff2019critical}, we find that critical exponents do not depend on the model of external noise. This suggests that this type of transition might be very generic, independent of the microscopic details in the underlying dynamics.

\section{Elasto-plastic models}
\subsection{Principle}

For this study we chose to use coarse-grained elasto-plastic models (EPMs), which provide a generic framework to describe the flow of soft glassy materials (for a review see Ref.~\cite{nicolas2018deformation} and references therein). 
In EPMs, the material first responds elastically to a global uniform driving (either by controlling the strain or the stress). The deformation or stress can then locally induce a rearrangement of particles \cite{argon1979plastic}, or plastic event, if a local threshold $\sigma_y$ is overcome. 
Such a plastic event causes a local relaxation of the stress and an elastic response of the surrounding material, and can trigger other local events.
The resulting macroscopic plastic flow is thus the consequence of these sequences of local rearrangements.
The particles rearrangements can be seen as plastic inclusions embedded in an elastic matrix, and are described as force quadrupoles (Eshelby problem). 
The elastic propagation kernel is thus described using the Eshelby propagator \cite{eshelby1957determination}, with an asymptotic power-law decay $\sim 1/r^d$ ($d$ being the spatial dimension) and a quadrupolar symmetry.

\subsection{Numerical model}
The spatially-resolved models considered in this work are extended from previous versions used to describe both steady-state flows of YSFs using a shear-imposed protocol model \cite{picard2005slow,nicolas2014rheology,martens2012spontaneous,liu2016driving},  and transient (creep) flow using a stress-imposed protocol model \cite{liu2018creep}.

We model an amorphous medium as a collection of mesoscopic blocks, each block being represented as a node ($i$,$j$) of a square lattice of size $L \times L$ (the lattice indices $i$,$j$ represent the discretised coordinates along $x$ and $y$ directions respectively).
The mesh size corresponds to the typical cluster size of rearranging particles undergoing a plastic rearrangement.
These local plastic transformations are assumed to have the same geometry as the globally applied simple shear, i.e., we consider a scalar model.
We decompose the total deformation of each node ($i$,$j$) into a local plastic strain $\gamma_{ij}^{pl}$, which is, in general, heterogeneous, and an elastic strain $\gamma_{ij}^{el}$. 

\subsubsection{Stress-imposed model}
When controlling the global stress in the system as in \cite{liu2018creep}, we also decompose the local stress into two parts, $\sigma_{ij}=\sigma^{\text{ext}}+\sigma_{ij}^{\text{int}}$, where $\sigma^{\text{ext}}$ is the externally applied uniform stress, and $\sigma_{ij}^{\text{int}}$ describes the stress fluctuations resulting from the elastic interactions between regions undergoing plastic deformation (i.e. particle rearrangements), as described by:
\begin{equation}
\sigma_{ij}^{\text{int}}=\mu\underset{i'j'}{\sum}G_{ij,i'j'}^{*}\gamma_{i'j'}^{pl} 
\end{equation}
The interaction kernel, $G^{*}$, is of Eshelby's type and is expressed, in Fourier space, as:
$\tilde{G^{*}}({\mathbf q}) = -4 \frac {q_x^2 q_y^2}{q^4}$ for $\mathbf{q} \neq 0$
and $\tilde{G^{*}}({\mathbf 0}) = 0$ 
so that $\sigma_{ij}^{\text{int}}$ describes the local stress fluctuations in a macroscopically stress-free state. 
Applying a macroscopic driving stress $\sigma^{\text{ext}}$ induces a uniform shift of the local stresses without altering internal fluctuations. 
The local dynamics is expressed as:
\begin{equation}
\frac{d}{dt}\gamma_{ij}^{pl}=n_{ij}\frac{\sigma_{ij}}{\mu\tau}=n_{ij}\frac{\sigma^{\text{ext}}+\sigma_{ij}^{\text{int}}}{\mu\tau}
\label{eq_stress_imposed}
\end{equation}
with $\mu$ the elastic modulus, $\tau$ a mechanical relaxation time setting the time units of the model and $\frac{d}{dt}\gamma_{ij}^{pl}=\frac{n_{ij}\sigma_{ij}}{\mu\tau}$ the strain rate produced by a plastic rearrangement occurring at a site $(i,j)$. The state variable $n_{ij}$, indicates whether the site  deforms plastically ($n_{ij}=1$) or elastically ($n_{ij}=0$), and has its own stochastic dynamics that will be described below.


\subsubsection{Strain-imposed model}
Another widely used model consists in imposing the strain rate to the system, as in \cite{picard2005slow,nicolas2014rheology,martens2012spontaneous,liu2016driving}. In this model, $\dot{\gamma}$ is the externally applied shear rate and we compute the local stress $\sigma_{ij}$, evolving with the over-damped dynamics:

\begin{equation}
\frac{d}{dt}\sigma_{ij}=\mu\dot{\gamma}+\mu\underset{i'j'}{\sum}G_{ij,i'j'}\frac{d}{dt}\gamma_{i'j'}^{pl}
\label{eq_strain_imposed}
\end{equation}
with $\frac{d \gamma_{ij}^{pl}}{dt} = \frac{n_{ij}\sigma_{ij}}{\mu \tau} $.
The interaction kernel $G$ is also of Eshelby's type \cite{eshelby1957determination}, and $\tilde{G}({\mathbf 0})\neq 0$ is determined by the integral over the whole system of the elastic response. It is related to $G^{*}$ by: $G=G^{*}-1/L^2$, because unlike $G^{*}$, $G$ does not describe the stress fluctuations in a macroscopically stress-free state, but the full stress field instead.

\subsubsection{Stochastic dynamics for the plastic activity}

Besides the dynamics described in either Eq.~(\ref{eq_stress_imposed}) or Eq.~(\ref{eq_strain_imposed}) depending upon the driving protocol, each node alternates between a local plastic state ($n_{ij}=1$) and a local elastic state ($n_{ij}=0$). 
The stochastic rules, as described in \cite{picard2005slow}, involve a rate of plastic activation $1/\tau_{pl}$ when the local stress exceeds a barrier $|\sigma_{ij}|>\sigma_{y}$ ($n_{ij}: 0 \rightarrow1 $)  and a rate $1/\tau_{el}$ for a plastic node turning elastic again ($n_{ij}: 1 \rightarrow 0$).
%
We consider in this work that a fluidising noise induces additional plastic events ($n_{ij}: 0 \rightarrow1 $) with a ``vibration rate'' $k_{vib}$. The three different types of transitions between elastic and plastic states are summarised below:

\begin{equation}
\begin{cases}
\begin{array}{ccc}
n_{ij}(t)\,: & 0\overset{1/\tau_{pl}}{\rightarrow}1 & \text{ if } \:\sigma_{i}>\sigma_{y}\\
n_{ij}(t)\,: & 0\overset{k_{vib}}{\rightarrow}1 & \forall\sigma_{i}\\
n_{ij}(t)\,: & 1\overset{1/\tau_{el}}{\rightarrow}0 
\end{array}
\end{cases}
\end{equation}
We study two different models for the activation of plastic events by an external noise:
\begin{enumerate}
\item Model 1: Constant activation rate: $k_{vib} = 1/\tau_{vib}$ for any value of the local stress $\sigma_{ij}$
\item Model 2: Arrhenius-like activation: $k_{vib} = k_0 e^{\lambda_{vib}(\sigma_{ij}-\sigma_y)}$ with $k_0$ a prefactor kept constant in our study and $\lambda_{vib}$ controlling the magnitude of the noise.
\end{enumerate}
These activated events have the same properties as the ones induced by shear, i.e., they lead to a redistribution of stress in the system through the Eshelby propagator.

In the following, the values of stress, strain rate and time are respectively given in units of $\sigma_{y}$, $\sigma_{y}/\mu\tau$ and $\tau$. 
We set $\tau_{pl}=1 $ and the restructuring time $\tau_{el}=10$ is chosen large compared to the other timescales in the system in order to induce local softening. Long restructuring times lead to non monotonic flow curves \cite{coussot2010physical} and are associated with permanent shear bands when imposing the shear rate in the system, as described in \cite{martens2012spontaneous}. 

We study the influence of an external noise by varying the value of the vibration rate $k_{vib}$, either varying $\tau_{vib}=k_{vib}^{-1}$ for model 1 (random activation), or $\lambda_{vib}$ for model 2 (Arrhenius-type activation) using both shear rate and stress controlled driving protocols, as they give access to different flow features in the case of non-monotonic flow curves. 
As we are interested in bulk quantities, we simulate the above elasto-plastic model using periodic boundary conditions in all directions.
Large scale simulations of the elasto-plastic model are performed using a GPU-based parallel implementation \cite{liu2016driving}.

\section{Average flow features}

\subsection{Rheology: flow curves}
\begin{figure}
\centering
\includegraphics[width=0.8\columnwidth, clip]{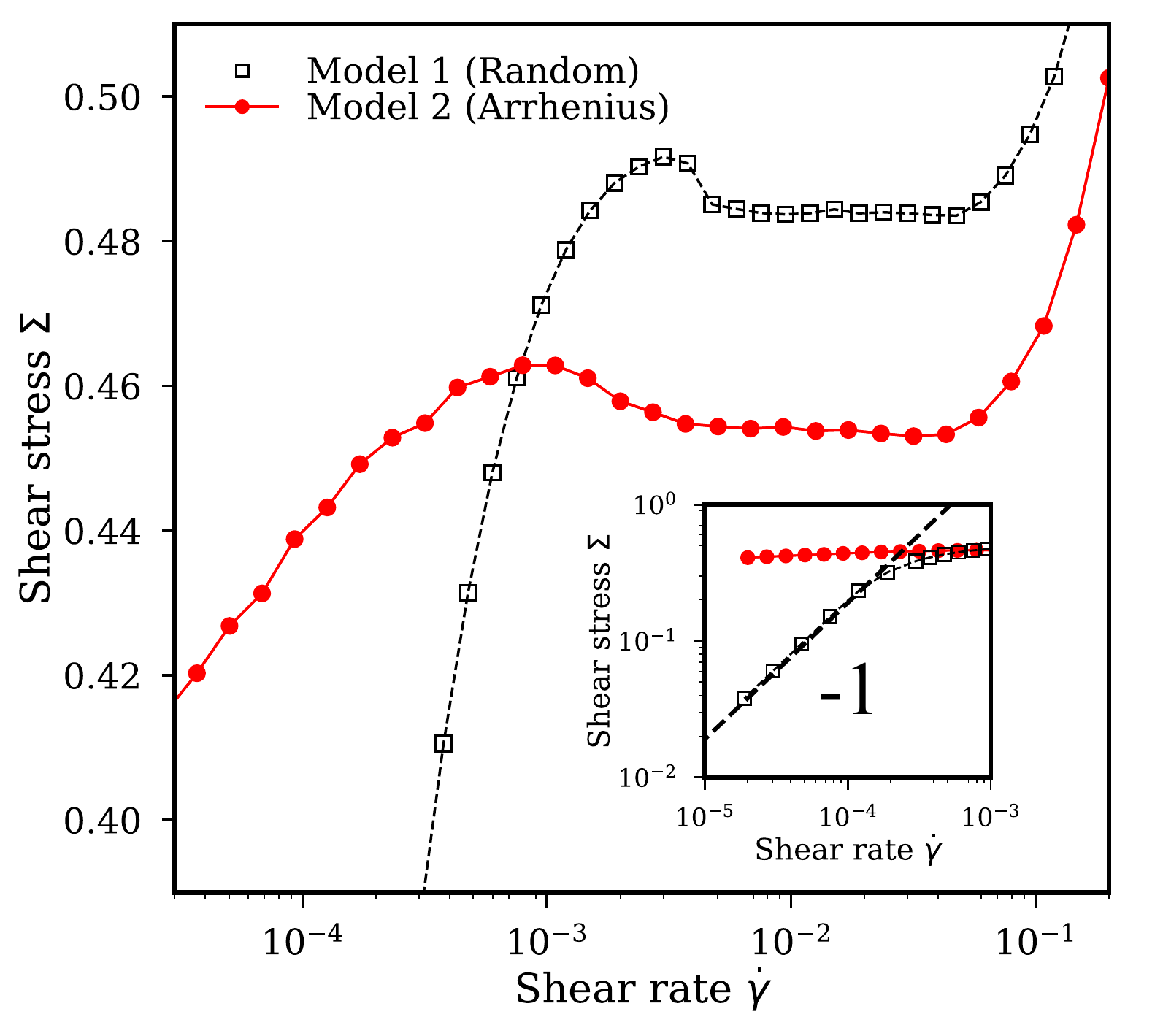}
\caption{Shear stress $\Sigma$ as a function of the imposed shear rate $\dot\gamma$ for the two models of agitation with $k_{vib}=3.3 \cdot 10^{-4}$ (model 1, black square symbols) and $\lambda_{vib}^{-1}=3.3 \cdot 10^{-2}$ (model 2, red dots), for a system size $L=256$. Inset: log-log plot showing the linear rheology regime at small stress for model 1.}.
\label{Fig2_Rheology}
\end{figure}

\begin{figure*}
\centering
\includegraphics[width=1.6\columnwidth, clip]{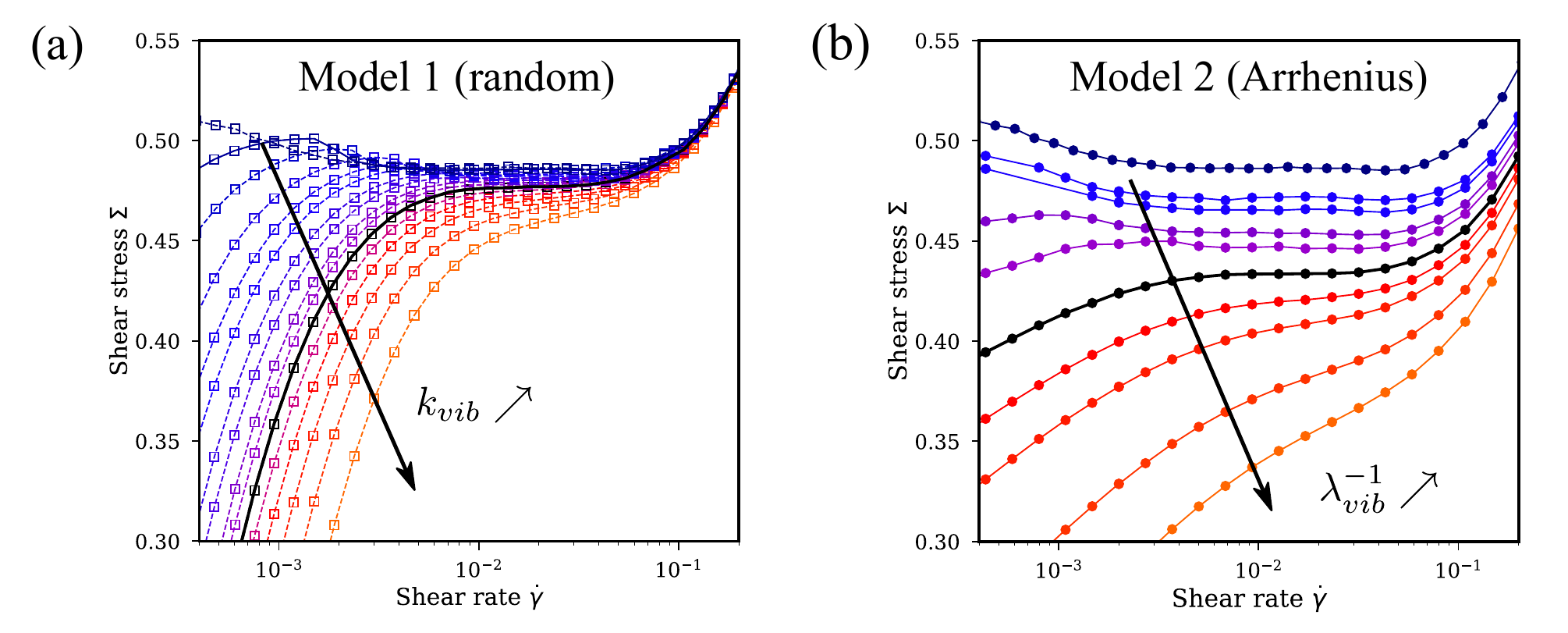}
\caption{Shear stress $\Sigma$ as a function of the imposed shear rate $\dot\gamma$ for various noise magnitude for (a) the random activation model (model 1, $k_{vib}$ ranging from $10^{-5}$ to $3.3 \cdot  10^{-3}$) and (b) the Arrhenius activation model (model 2, $\lambda_{vib}^{-1}$ ranging from $2 \cdot 10^{-2}$ to $10^{-1}$), for a system size $L=256$. The upper curve in (a) and (b) is obtained in absence of noise.}
\label{Fig2_Phenomenology}
\end{figure*}

\begin{figure*}
\centering
\includegraphics[width=1.6\columnwidth, clip]{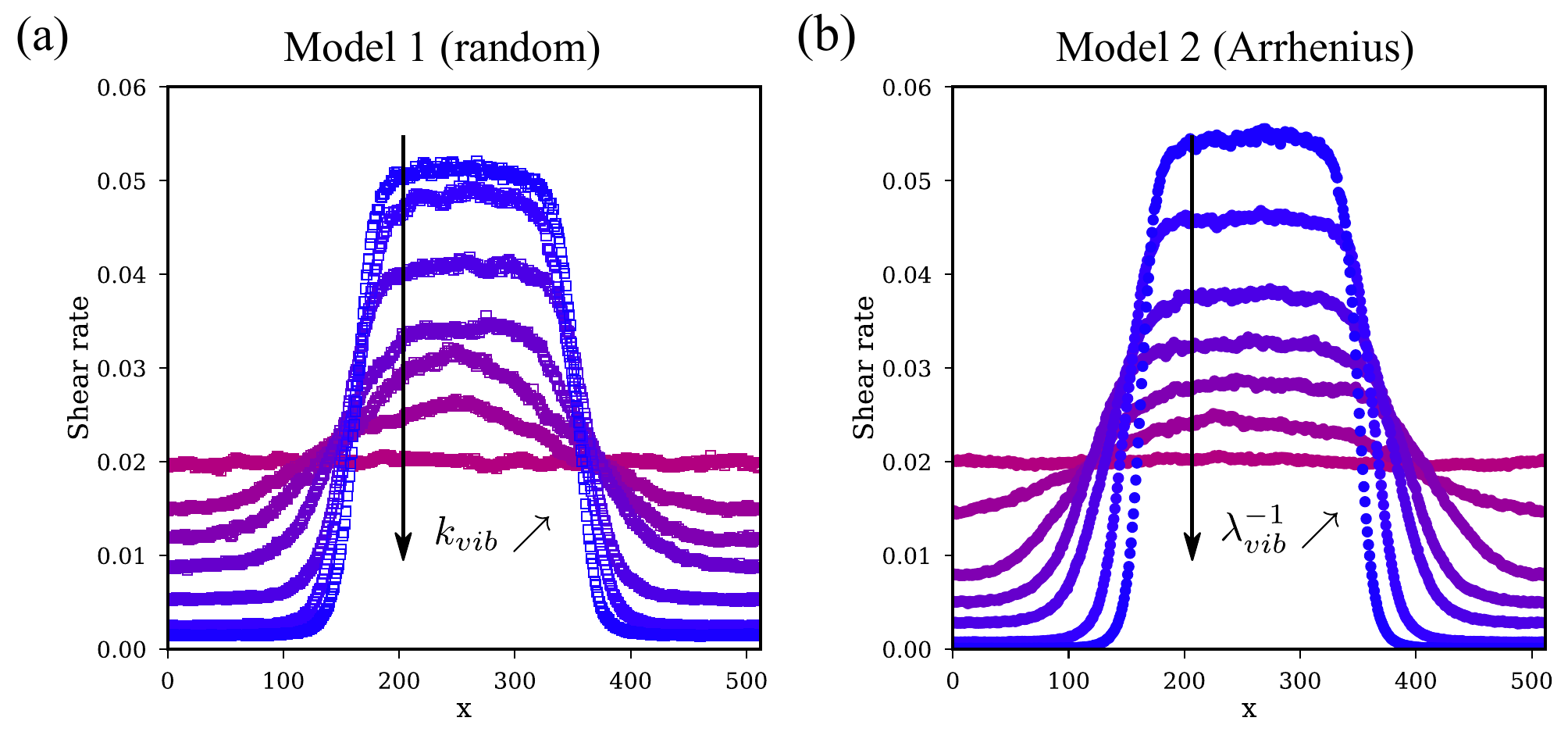}
\caption{Steady state shear rate profiles (computed over a strain window of 1000) for various values of the noise (for a system size $L=512$) for (a) the random activation model ($k_{vib}^{-1}$ ranging from $3.3 \cdot 10^{-4}$ to $1.4 \cdot 10^{-3}$) and (b) for the Arrhenius activation model ($\lambda_{vib}^{-1}$ ranging from $2.5 \cdot 10^{-2}$ to $5 \cdot 10^{-2}$).}.
\label{Fig3_Shear_Bands}
\end{figure*}

We compute the average steady state shear stress as a function of the imposed shear rate (flow curve) for the two models of noise (Fig.~\ref{Fig2_Rheology}), with magnitudes  $k_{vib}=3.3 \cdot 10^{-4}$ for model 1 (random activation) and $\lambda_{vib}^{-1}=3.3 \cdot 10^{-2}$ for model 2 (Arrhenius) respectively. 
Let us first describe the generic features of the flow curves, irrespective of the model of noise.
The system exhibits a fluid-like behaviour at low stress (and $\dot\gamma < 10^{-3}$), while in the absence of noise, the elasto-plastic model considered in this work exhibits a finite yield stress at $\dot\gamma \rightarrow 0$ (upper dark blue curves in Fig.~\ref{Fig2_Phenomenology}).
This regime is followed by a stress-plateau ($10^{-3}<\dot\gamma < 10^{-1}$), associated with a shear banding instability (see Fig.~\ref{Fig3_Shear_Bands}). 
The decreasing part of the flow curve around $\dot\gamma = 10^{-3}$, reminiscent of the non-monotonic constitutive flow curve, is due to finite size effects and will not be discussed in detail here. The last regime ($\dot\gamma > 10^{-1}$) corresponds to a stable homogeneous flow.


While the global shape of the flow curve is the same, the rheological behaviour in the low stress regime differs between the two models of activation. 
The random activation rule (model 1) leads to a Newtonian behaviour (as shown in the log-log plot of the inset of Fig.~\ref{Fig2_Rheology}), whereas the Arrhenius-like rule (model 2) leads to a logarithmic-like flow behaviour, reminiscent of experiments on vibrated granular media (see Fig.~1 in the work by Wortel et al.~\cite{wortel2016criticality}).
This can be understood simply from the activation rule. 
Assuming that the shear rate is proportional to the stress multiplied by the average number of active sites $\dot \gamma \propto \Sigma \langle n \rangle $ (which is reasonable in the case of a homogeneous flow \cite{martens2012spontaneous})  and that, in the low stress regime, the mean number of active sites is proportional to the vibration rate $k_{vib}$, then  $\dot \gamma \propto \Sigma k_{vib} $. This leads to a linear rheological behaviour for model 1:  $ \Sigma \propto \dot \gamma  \tau_{vib}$, and a non linear rheology for model 2, with $k_{vib} = k_0 e^{\lambda_{vib}(\sigma_{ij}-\sigma_y)}$.


Flow curves for various noise magnitudes are depicted in Fig.~\ref{Fig2_Phenomenology} for the two models of activation
(by varying either $k_{vib}=1/\tau_{vib}$ in model 1 or $\lambda_{vib}$ in model 2).
For the two models, the effect of noise is (i) a fluidisation (vanishing yield stress) at any value of the noise magnitude and (ii) a transition from a non-monotonic to a monotonic flow curve at a noise magnitude, $k_{vib}^c = (1.3 \pm 0.2) \cdot 10^{-3}$ or $\lambda_{vib}^c = 20 \pm 2$ (the thick black lines in  Fig.~\ref{Fig2_Phenomenology} correspond to $k_{vib}^c = 1.25 \cdot 10^{-3}$ and $\lambda_{vib}^c = 20$).


\subsection{Shear rate profiles}

Fig.~\ref{Fig3_Shear_Bands} depicts profiles of shear rate in steady state averaged over a strain window of 1000, for the two models of noise.
In the low noise regime (non-monotonic flow curves) the system separates into two flowing regions (blue curves in Fig.~\ref{Fig3_Shear_Bands}), where the minimum and the maximum of the shear rate profile are determined by the boundaries of the stress plateau in Fig.~\ref{Fig2_Phenomenology}.
This is similar to the shear bands reported by Martens et al. \cite{martens2012spontaneous} within a mesoscale elasto-plastic model in the absence of fluidising noise, leading to the coexistence of a flowing and an arrested region instead. 
In our model, the difference in shear rate between the two bands decreases  as the magnitude of the noise is increased, until reaching a stable homogeneous flow regime as shown by the flat profiles (light purple curves in Fig.~\ref{Fig3_Shear_Bands}) corresponding to a monotonic flow curve.

The transition from a phase separated flow to a homogeneous flow can thus be characterised using the difference in shear rate between the two flowing bands.
We define the order parameter of this transition as the logarithm of the ratio of shear rates in the two flowing bands: 

\begin{equation}
\text{Order parameter} = \log (\frac{\dot\gamma_{fast}}{\dot\gamma_{slow}}) = S_{fast}-S_{slow} 
\end{equation}

\noindent with $S = \log(\dot\gamma)$.

Note that this transition can also be characterised using a stress-controlled protocol \cite{legoff2019critical}. When imposing the stress to the system, the flow is necessarily homogeneous and hence the unstable (negative slope) part of the constitutive flow curve becomes inaccessible. By varying the initial conditions for the flow, the width of this unstable region can thus be characterised, giving access to the extent of the stress plateau, as done indirectly here by studying the shear rate profiles. As a consistency check, we will see in section \ref{sec:Critical_point} that the scaling of the order parameter near the transition is independent of the protocol used. 

\begin{figure}
\centering
\includegraphics[width=0.8\columnwidth, clip]{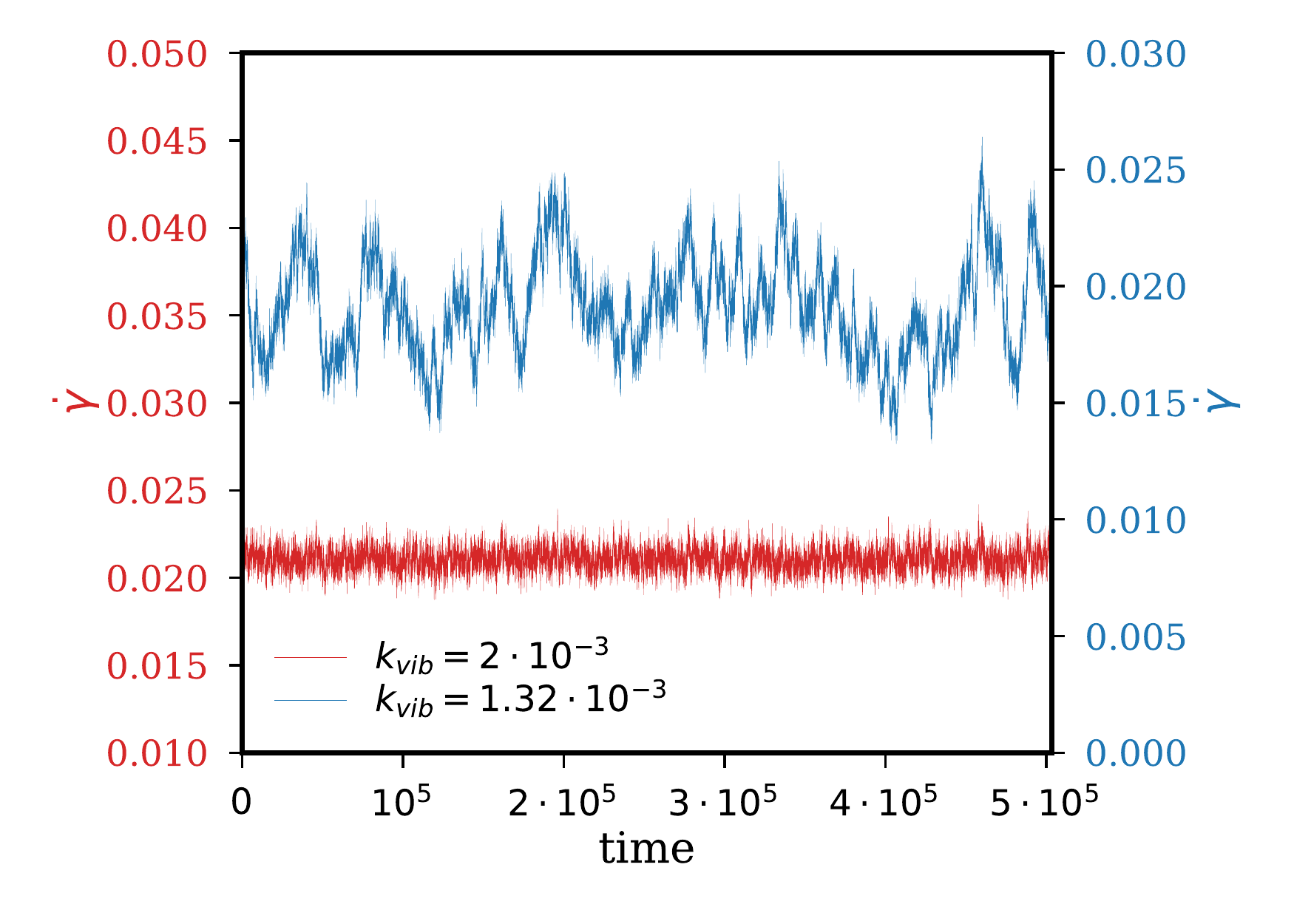}
\caption{Macroscopic shear rate $\dot\gamma$ as a function of time for $k_{vib}=1.325 \cdot 10^{-3}$ (top, blue curve) and  $k_{vib}=5 \cdot 10^{-3}$ (bottom, red curve) for a system size $L=512$ (the $y$-axes of the two curves are shifted for readability).}
\label{Fig_example_fluctuations}
\end{figure}

\section{\label{sec:Giant_fluctuations} Giant shear-rate fluctuations}

\begin{figure*}
\centering
\includegraphics[width=1.6\columnwidth, clip]{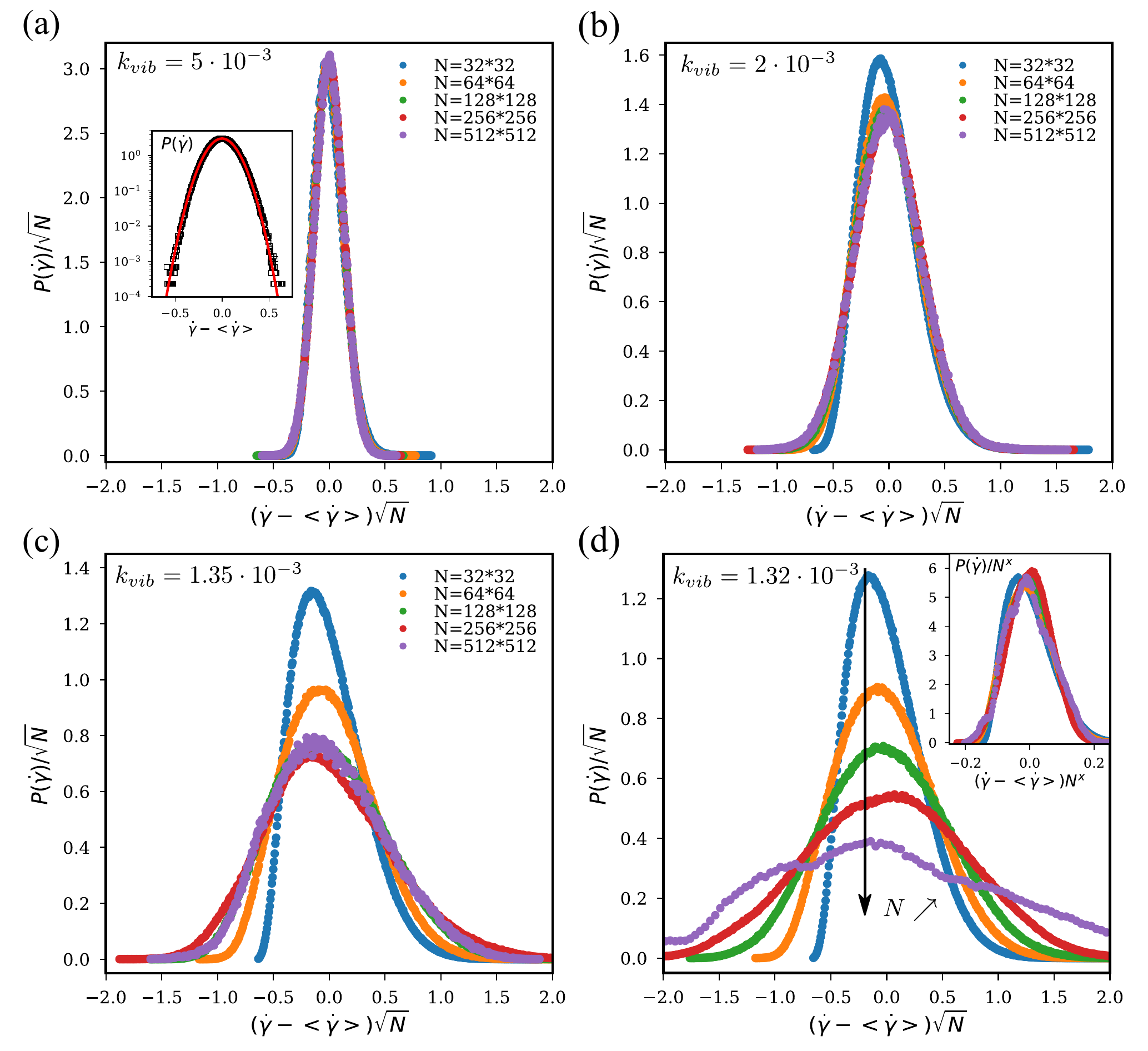}
\caption{Centred distributions of macroscopic shear rate $\dot\gamma$ computed for model 1 (random activation) at the inflexion point of the flow curve for various system sizes (from $L=32$ to $L=512$), with the $x$-axis rescaled by multiplying $\dot\gamma$ by the linear system size $L=\sqrt{N}$ and the $y$-axis by dividing $P(\dot\gamma)$ by $\sqrt{N}$, in the stable flow phase (monotonic flow curve), for decreasing values of the noise magnitude: (a) $k_{vib}=5 \cdot 10^{-3}$, (b) $k_{vib}=2 \cdot 10^{-3}$, (c) $k_{vib}=1.35 \cdot 10^{-3}$ and (d) $k_{vib}=1.32 \cdot 10^{-3}$. Inset of (a): Lin-log plot of the distribution for $L=256$ with Gaussian fit (red). Inset of (d): Finite size data collapse of the shear rate distributions using an exponent $x=0.275$.}
\label{Fig_giant_fluctuations}
\end{figure*}

We now investigate in more details the high noise regime corresponding to $k_{vib}>k_{vib}^c$ (model 1) or $\lambda_{vib}^{-1}>{\lambda_{vib}^c}^{-1}$ (model 2), where the flow curve is monotonously increasing, so that the homogeneous flow is stable. 
In this regime, using shear- or stress-controlled protocols leads to the same average rheological behaviour.
Interesting properties are rather to be found in shear-rate fluctuations (using a stress-controlled protocol), that we first describe qualitatively.

We measure the macroscopic shear rate in the system as a function of time in steady state, for two different values of $k_{vib}$ (for model 1)[Fig.~\ref{Fig_example_fluctuations}], choosing stress values corresponding to the inflexion point of the flow curve.
For large values of the noise magnitude ($k_{vib}=5 \cdot 10^{-3}$, lower red curve in Fig.~\ref{Fig_example_fluctuations}), the fluctuations of shear rate $\dot\gamma$ are relatively small (variations of about 10 \text{\%} of the mean value for a system size $N=512^2$) and not correlated in time. 
When decreasing the noise towards its value at the transition between monotonic and non-monotonic flow curves ($k_{vib}= 1.32 \cdot 10^{-3}$, see Fig.~\ref{Fig2_Phenomenology}), the fluctuations of $\dot\gamma$ increase (50 \text{\%}) and become correlated in time, as it can be seen on the upper (blue) curve of Fig.~\ref{Fig_example_fluctuations}.

\subsection{Rescaled shear-rate distributions}
To perform a quantitative analysis, we compute the distributions of macroscopic shear rate in steady state for various system sizes $N$ and noise magnitudes $k_{vib}$, again for a stress value corresponding to the inflexion point of the flow curves.
From the central limit theorem, one expects relative fluctuations of the shear rate to scale as $1/\sqrt{N}$ for large system size.
Fig.~\ref{Fig_giant_fluctuations} depicts the centred distributions of $\dot\gamma$ rescaled by $\sqrt N$ for different system sizes $N$.
In this representation, curves collapse if relative fluctuations scale like $1/\sqrt{N}$.
For large noise magnitudes (Fig.~\ref{Fig_giant_fluctuations}(a), $k_{vib}=5 \cdot 10^{-3}$), the data for all system sizes collapse onto the same curve, indicating that the shear rate fluctuations obey the central limit theorem.
Unsurprisingly, the shear rate fluctuations in this regime follow a Gaussian distribution as shown by the fit in the inset of Fig.~\ref{Fig_giant_fluctuations}(a).
As the noise magnitude is decreased (Fig.~\ref{Fig_giant_fluctuations}(b-d)), the rescaled distributions widen and a systematic dependence with the system size appears.
This indicates a deviation from the central limit theorem at the approach of the transition, associated with growing spatial correlations of the macroscopic shear rate.
Moreover, the maximum system sizes for which finite size effects are observed in Fig.~\ref{Fig_giant_fluctuations} give an estimate of the correlation length $\xi$ in the system as the noise is varied (in Fig.~\ref{Fig_giant_fluctuations}(b) $32<\xi<64$ for $k_{vib}=2 \cdot 10^{-3}$, (c) $64<\xi<128$ for $k_{vib}=1.35 \cdot 10^{-3}$ and (d) $\xi>512$ for $k_{vib}=1.32 \cdot 10^{-3}$).
The increase of the correlation length when decreasing the noise indicates a possibly diverging length scale in the system at the transition, which is consistent with the existence of a critical point \cite{wortel2016criticality, legoff2019critical}. 
We show in the inset of Fig.~\ref{Fig_giant_fluctuations}(d) that the distribution for $k_{vib}=1.32 \cdot 10^{-3}$ can be approximately collapsed by rescaling the shear rate by $N^x$ with $x \simeq 0.275$ (instead of $1/2$ far from the transition). This shows that relative fluctuations of the shear rate decay approximately as $1/N^{0.275}$ with system size, that is, much more slowly than the standard $1/\sqrt{N}$ scaling corresponding to the central limit theorem. This slower decay of relative fluctuations with system size can be described as the presence of giant shear-rate fluctuations.
These giant shear-rate fluctuations are directly visible in an experimental context \cite{wortel2016criticality}, and their characterisation is thus of interest.

\begin{figure*}
\centering
\includegraphics[width=1.6\columnwidth, clip]{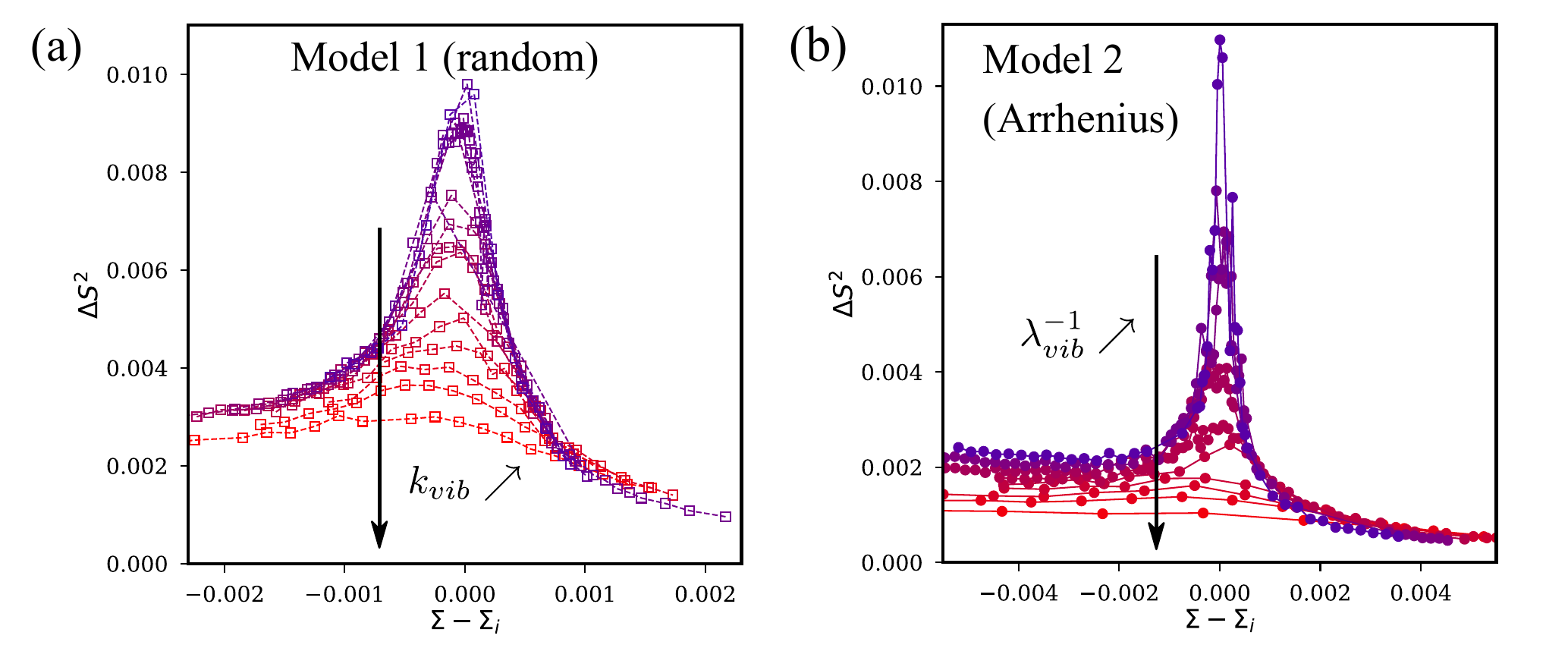}
\caption{Variance of $S=\log(\dot\gamma)$, $\Delta S ^2$, as a function of the distance (in stress) to the inflexion point of the flow curve, $\Sigma - \Sigma_i$ for various noise magnitudes and for a system size $L=128$. (a) Model 1, for values of $k_{vib}$ ranging from $1.29 \cdot 10^{-3}$ (upper purple curve) to $1.75 \cdot 10^{-3}$ (lower red curve). (b) Model 2, for values of $\lambda_{vib}^{-1}$ ranging from $5.12 \cdot 10^{-2}$ to $6.6 \cdot 10^{-1}$. }
\label{Fig6_fluctuations_stress}
\end{figure*}

Note that for a linear system size $L$ ($=\sqrt{N}$) much larger than the correlation length of the system, the standard $1/\sqrt{N}$ scaling is recovered, together with a Gaussian shape of the distribution. This crossover is visible on Fig.~\ref{Fig_giant_fluctuations}(b) and (c).

Instead of looking at the dependence of fluctuations on system size, one may also look at their dependence on stress.
In Fig.~\ref{Fig6_fluctuations_stress}, we depict the variance of $S$ as a function of the distance (in stress) to the inflexion point of the flow curve for various noise magnitudes (for a single system size, $L=128$). It can be clearly seen that the variance at the inflexion point is maximal (for $\Sigma = \Sigma_i$) and increases monotonously as the noise magnitude is decreased towards its value at the transition. The same phenomenology is observed for model 2 (Fig.~\ref{Fig6_fluctuations_stress}(b)). 
Enhanced fluctuations are thus observed close to the inflexion point, even for a fixed, intermediate system size.

\subsection{Origin of non-standard fluctuations}
The reason for the emergence of fluctuations with non-trivial statistics is actually the presence of a critical point, around which the system becomes correlated over a large length scale.
Quite interestingly, non-trivial scalings are found close to the critical point not only for fluctuations, but also for average quantities, which may make some of the critical properties easier to measure.
Both average quantities and fluctuations exhibit power-law scalings at the transition, and the critical point is described by a set of critical exponents, which are supposed not to depend on details of the system, but rather only on generic properties shared by a broad class of systems. For instance, one may expect that both model 1 and model 2 share the same critical properties. This statement, though, does not rest on any firm theoretical consideration, and should be tested numerically. This is one of the goals of the next section.
In a previous work \cite{legoff2019critical}, we characterised in detail this critical point for the random activation model (model 1).
In the next section, we perform a similar analysis for the Arrhenius activation model (model 2), and show that the scaling of both average quantities and fluctuations does not depend on the model of noise in the critical regime.

\section{\label{sec:Critical_point} Generic critical point at finite shear and vibration rates}

We first study the evolution of average steady state quantities as the noise magnitude is varied in the critical regime by studying the scaling of the flow curves and of the order parameter.
Then, we investigate the scaling of shear rate fluctuations.

In the following, the noise magnitude is designated by the relative distance to the critical point, which reads, for the two models of noise:

\begin{eqnarray}
    \varepsilon &=& \frac{k_{vib}-k_{vib}^c}{k_{vib}^c} \qquad \; \text{ (Model 1) } \\
 \varepsilon &=& \frac{\lambda_{vib}^{-1}-{\lambda_{vib}^c}^{-1}}{{\lambda_{vib}^c}^{-1}} \quad \, \text{ (Model 2) }
\end{eqnarray}

The stable flow (noise dominated) regime thus corresponds to $\varepsilon > 0$ and the phase separated regime to $\varepsilon < 0$. We summarise below the list of scalings expected at the transition.

In the regime $\varepsilon < 0$, the order parameter $S_{fast} -  S_{slow}$ vanishes as the noise magnitude is increased towards its critical value as a power law with an exponent $\beta$:
\begin{equation}
    S_{fast} -  S_{slow} \sim | \varepsilon | ^\beta \text{ for } \varepsilon<0
    \label{eq_op_beta}
\end{equation}
The critical point corresponds to the transition from a monotonic to a non-monotonic flow curve, and hence the slope of the flow curve at the inflexion point vanishes at the critical point. It can be interpreted as an inverse susceptibility $\chi$, expected to scale as a power law of the distance to the critical point $\varepsilon$ with an exponent $\gamma$:
\begin{equation}
    \chi \sim  \varepsilon  ^{-\gamma} \text{ for } \varepsilon>0
    \label{eq_chi_gamma}
\end{equation}
At the critical point ($\varepsilon=0$), the flow curve exhibits a zero slope and $S$ is expected to vary as a power law of the imposed shear stress (after centring the flow curve using the coordinates of the critical point ($S_c,\Sigma_c$)) with an exponent $1/\delta$:
\begin{equation}
    |S - S_c| \sim | \Sigma - \Sigma_c | ^{1/\delta} \text{ for } \varepsilon=0
    \label{eq_Fc_delta}
\end{equation}

The variance of $S$ diverges at the critical point when varying the imposed stress:
\begin{equation}
    \Delta S^2 \sim | \Sigma - \Sigma_c | ^{-\kappa} \text{ for } \varepsilon=0
    \label{eq_Fluc_kappa}
\end{equation}

The variance and the correlation time of $S$ diverge as the critical point is approached from the high noise regime:
\begin{eqnarray}
\Delta S^2 \sim  \varepsilon  ^{-\gamma*} \text{ for } \varepsilon>0 \\
\tau_{corr} \sim  \varepsilon  ^{-\mu} \text{ for } \varepsilon>0 
\end{eqnarray}

\subsection{Scaling of average quantities}

\paragraph* {Scaling of the flow curves and susceptibility:}
Using a stress-controlled protocol, we compute the steady state flow curve in the stable flow regime ($\varepsilon > 0$), for the two models of fluidising noise and investigate the scaling of the stress $\Sigma$ as a function of $S=\log(\dot\gamma)$.
The data for the two models can be well fitted to a Landau type expansion in the critical regime:
\begin{equation}
\Sigma=\Sigma_{i}+a\left(S-S_{i}\right)^{3}+b\left(S-S_{i}\right)
\label{fit_Landau}
\end{equation}
where $a$, $b$, $S_i$ and $\Sigma_i$ are fitting parameters shown in Appendix C (Fig.~\ref{fig_fitting_params}) for various values of $\varepsilon$ and system sizes $L$. 

\begin{figure}
\centering
\includegraphics[width=0.8\columnwidth, clip]{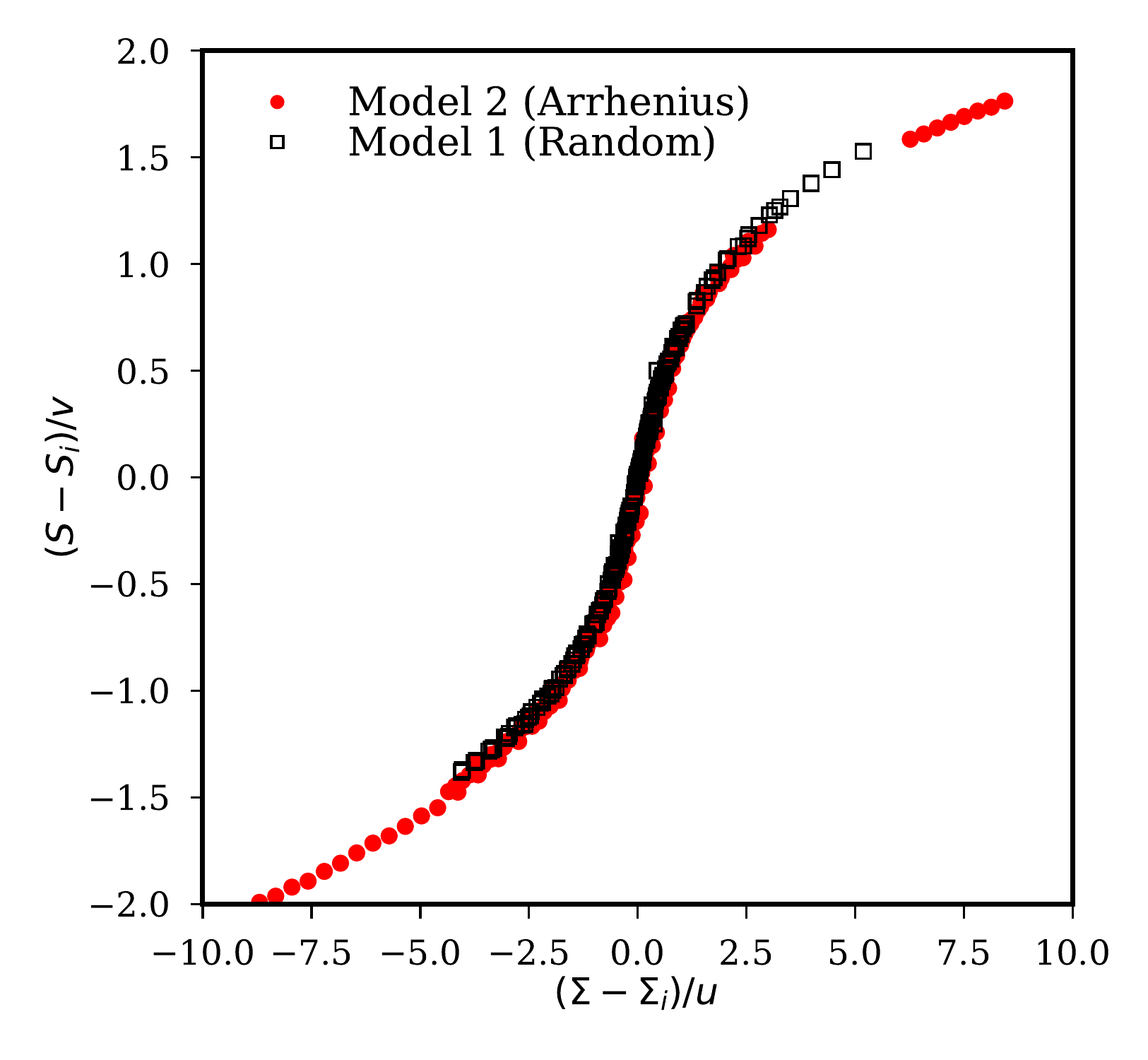}
\caption{Data collapse for the two models using the fitting parameters from the Landau expansion fit of the flow curve ($a(\varepsilon)$, $b(\varepsilon)$, $S_i(\varepsilon)$ and $\Sigma_i(\varepsilon)$) with $u=\sqrt{b/a}$ and $v=\sqrt{b^3/a}$. Data shown for a system size $L=128$.}
\label{fig_flow_curve_fit}
\end{figure}
Using these fitting parameters, the data from the two models can be collapsed onto the same master curve, as shown in Fig.~\ref{fig_flow_curve_fit}.
$a(\varepsilon)$ is roughly constant (Fig.~\ref{fig_fitting_params}(a)).
The coordinates of the inflexion point ($S_i(\varepsilon)$,$\Sigma_i(\varepsilon)$) evolve monotonously as the noise is varied, and describe the analogous of the super-critical liquid-gas boundary in equilibrium phase transitions (Fig.~\ref{fig_fitting_params}(c-d)).  
The prefactor $b$ of the linear term in Eq.~(\ref{fit_Landau}), (Fig.~\ref{fig_fitting_params}(b)), decreases linearly with $\varepsilon$ and vanishes at the critical point. As it describes the slope of the flow curve at the inflexion point, it is interpreted as the inverse susceptibility $b=1/ \chi$, which diverges at the critical point with an exponent $\gamma \simeq 1$ [Eq.~(\ref{eq_chi_gamma})] (see Fig.~\ref{fig_FFS_fluctuations_2models}).
At the critical point, $b=0$ and hence the flow curve (``critical isotherm'') is well described by Eq.~(\ref{eq_Fc_delta}), with $\delta = 3$.

\paragraph*{Critical point location:}
To locate the critical value of noise, we fit the data of Fig.~\ref{fig_fitting_params}(b) to extract the value of $k_{vib}^c$ and $\lambda_{vib}^c$ for which $b = 0$ (diverging susceptibility).
While significant finite size effects are observed for model 1 (random) (and discussed in detail in \cite{legoff2019critical}), the finite size effects on the value of $b$ remain within the error bars for model 2.
We find $k_{vib}^c = (1.35 \pm 0.01) 10^{-3}$ for model 1 \cite{legoff2019critical} and ${\lambda_{vib}^c}^{-1} = (5.1 \pm 0.2) 10^{-2}$ for model 2. 

\begin{figure}
\centering
\includegraphics[width=0.8\columnwidth, clip]{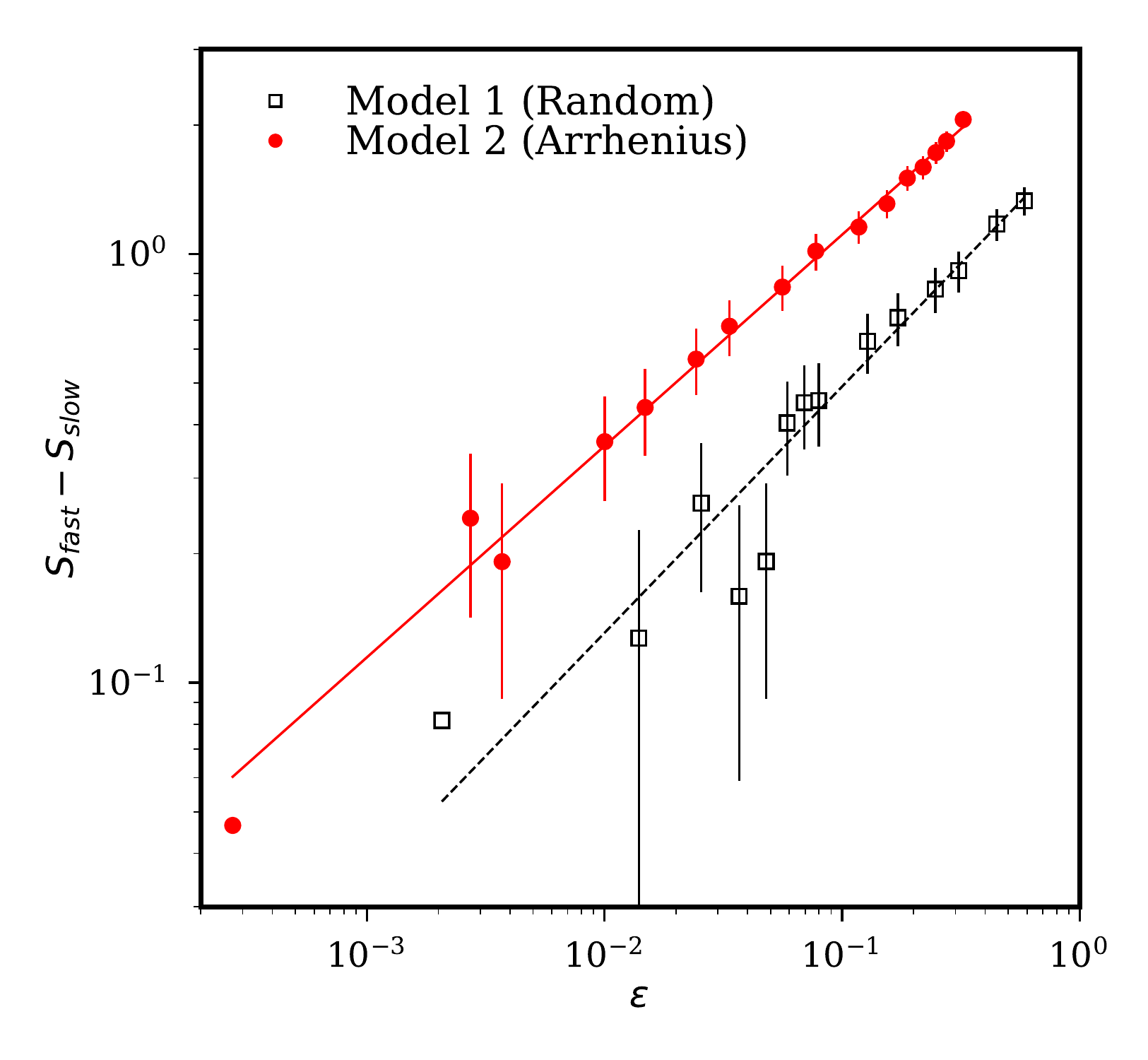}
\caption{Order parameter $S_{fast} -  S_{slow}$ as a function of the scaled distance to the critical noise magnitude $\varepsilon$, $\lambda_{vib}^c$ and $k_{vib}^c$  being  fitting parameters of the fit $S_{fast} -  S_{slow}= A |\varepsilon|^{\beta}$ (see text). Model 1: lower black square symbols (with dashed line fit). Model 2:  Upper solid circles (with solid line fit). Error bars are estimated from the standard deviation of $S$ in the bands.}
\label{fig_OrderParam}
\end{figure}

\paragraph*{Order parameter:} In Fig.~\ref{fig_OrderParam}, we investigate the scaling of the order parameter $S_{fast}-S_{slow}$ extracted from the shear rate profiles (Fig.~\ref{Fig3_Shear_Bands}) at the approach of the transition, in the regime $\varepsilon<0$. The order parameter decreases as $\varepsilon$ is decreased, and scales as a power law of the distance to the critical point (fit of the form $S_{fast} -  S_{slow}= A |\varepsilon|^{\beta}$).
In the fit of Fig.~\ref{fig_OrderParam}, $\varepsilon = (k_{vib}-k_{vib}^c)/k_{vib}^c$ for model 1, $\varepsilon = (\lambda_{vib}-\lambda_{vib}^c)/\lambda_{vib}^c$ for model 2 and $k_{vib}^c$, $\lambda_{vib}^c$, 
$\beta$ and $A$ are free fitting parameters.
We get $\beta_1 = 0.58 \pm 0.07$ (model 1) and $\beta_2 = 0.49 \pm 0.04$ (model 2), with $k_{vib}^c = (1.21 \pm 0.05)\cdot 10^{-3}$ and $({\lambda_{vib}^c}^{-1} = 4.95 \pm 0.05)\cdot 10^{-2}$. 
For the two models, the critical exponent $\beta$ is close to 0.5, which is consistent with the Landau-type scaling in the regime $\varepsilon>0$. 

The critical noise magnitudes ($k_{vib}^c$, ${\lambda_{vib}^c}^{-1}$) obtained from the fit for a system size $L=512$, although they are not far from the previous estimate, are slightly underestimated.
In fact, extracting the order parameter from the flow profiles is a difficult task near the critical point, as the coarsening time of the shear bands increases (and diverges for $\varepsilon=0$).
As a consequence, we cannot access steady state profiles in the critical regime.
The configurations associated with the closest data points to the critical point in Fig.~\ref{fig_OrderParam} are not coarsened yet, thus leading to large error bars and possibly explaining the slight underestimate of $\Delta S$ and hence an underestimate of the critical noise magnitudes.
Using a stress-controlled protocol in our previous work \cite{legoff2019critical}, we were able to get accurate data close to the critical point and confirm the above scaling for model 1 (random), with $\beta = 0.52 \pm 0.02$.

\begin{figure}
\centering
\includegraphics[width=0.8\columnwidth, clip]{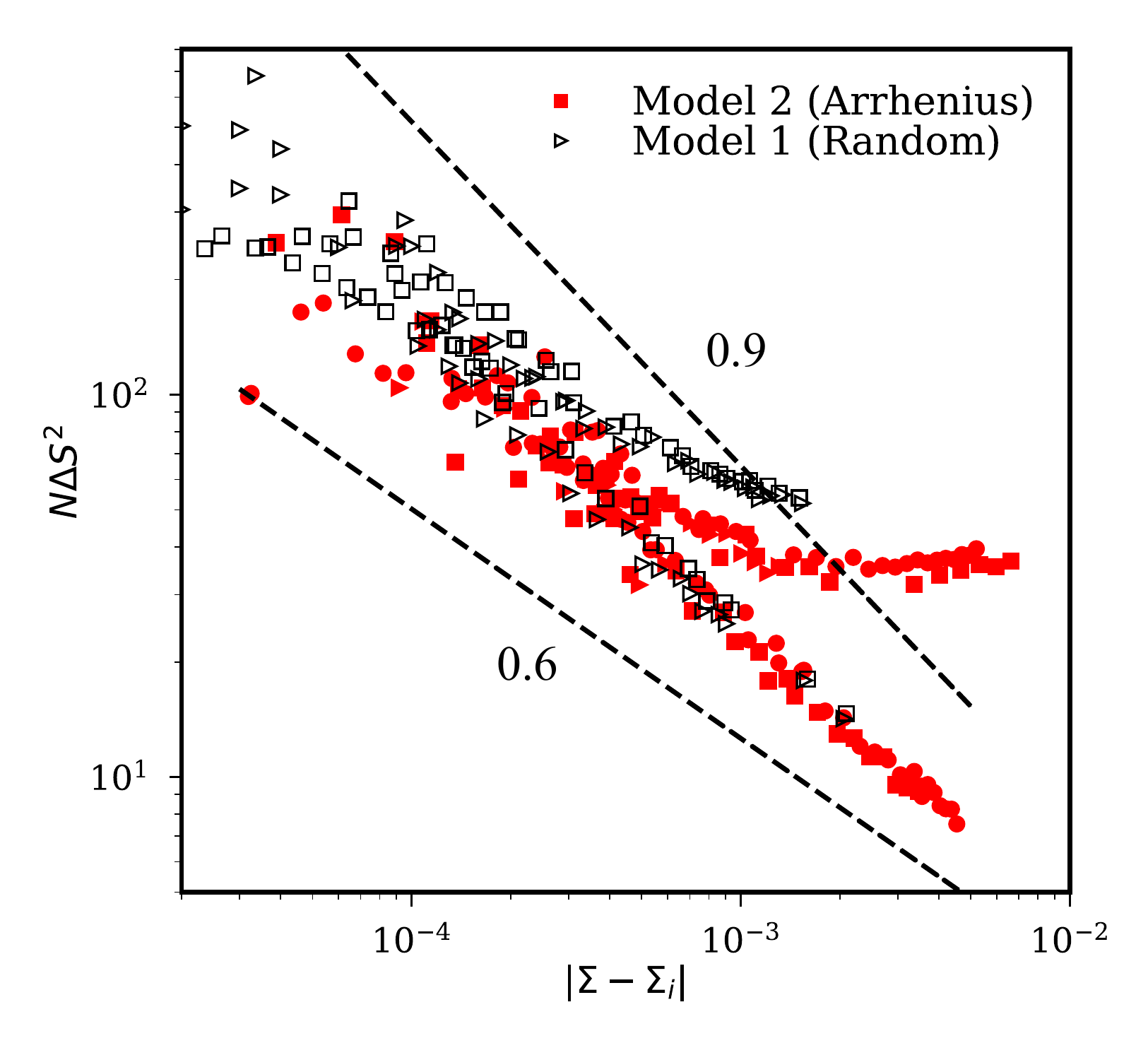}
\caption{Variance of $S$ multiplied by the number of sites, $N\Delta S ^2$ as a function of $|\Sigma-\Sigma_i|$ for values of the noise magnitude close to the critical value ($\varepsilon \simeq 0$).
Data for model 1 (black empty data points) and model 2 (red solid points) and various system sizes (circles: $L=128$, squares: $L=256$, triangles: $L=512$). 
Dashed lines: power-law guides to the eye with exponents $0.6$ and $0.9$.}
\label{fig_fluctuations_sigma}
\end{figure}

\paragraph*{Finite size effects:}
Let us now recall some of the main results of a former finite size data analysis that we performed for the random activation model (model 1) \cite{legoff2019critical}, and that we will use to perform the finite size data collapse in this section.
At the critical point, the correlation length of the system diverges as: 
\begin{equation}
    \zeta \sim \varepsilon^{-\nu} \text{ for } \varepsilon>0
\end{equation}
Formally, we performed direct measurements of $\zeta$ for model 1 \cite{legoff2019critical} and found $\nu \in [0.9,1]$. This was consistent with the finite size shift of the transition, which yielded a value of $\nu \simeq 1$.  

Using $\nu = 1$, we show in Fig.~\ref{fig_FFS_fluctuations_2models}(a) the data for the susceptibility $\chi = 1/b$ rescaled by a factor $L^{-\gamma/\nu}$ as a function of the scaled distance to the critical point $\varepsilon L^{1/\nu}$. 
$\varepsilon$ is expressed using the above value $k_{vib}^c = (1.35 \pm 0.01) 10^{-3}$ and refining the value of ${\lambda_{vib}^c}^{-1}$ to get the best data collapse (${\lambda_{vib}^c}^{-1} = 5.15 \cdot 10^{-2}$).
The ability to collapse the data for all system sizes for the two models of noise suggests that they share similar critical exponents for the susceptibility and the correlation length ($\gamma \simeq 1$ and $\nu \simeq 1$).

\subsection{Scaling of fluctuations}

In equilibrium systems, fluctuations are related to average quantities like the susceptibility, and at a critical point, the divergences of both quantities are related.
In this part we investigate the scaling of the fluctuations of $S$ when varying the imposed stress and the noise magnitude in the critical regime. 

\paragraph*{Varying the imposed stress in the critical regime:}

As shown in Fig.~\ref{Fig6_fluctuations_stress}, the variance of the order parameter increases as the stress approaches its value at the inflexion point of the flow curve, where it is maximal.
In figure \ref{fig_fluctuations_sigma}, we report, in a log-log plot, for the value of the noise closest to the critical point ($\varepsilon \simeq 0$), the value of the variance of $S$ as a function of the distance to the inflexion point, for various system sizes.
Let us point out that the scattering of data near the maximum is due to finite size effects, where the smallest values correspond to the smallest systems.
We find that the variance of $S$ varies as a power law of the distance to the critical stress, with exponents $\kappa_1 = 0.82 \pm 0.12$ (model 1) and $\kappa_2 = 0.73 \pm 0.15$ (model 2), that do not depend significantly on the noise model.
From the Landau fit of Eq.~(\ref{fit_Landau}), the susceptibility  ($\chi = \partial S /\partial \Sigma$ for $\varepsilon = 0$) varies as ($|\Sigma-\Sigma_c|^{2/3}$), which is within the error bars of our estimate of $\kappa$. 
Note that a similar scaling is found for the correlation time of the fluctuations (see Appendix B, Fig.~\ref{fig:Appendix_corr}(b)).

While the extent of the power law spans two decades in the ``shear dominated" regime (for $\Sigma>\Sigma_i$, lower data points in Fig.~\ref{fig_fluctuations_sigma}(c)), the range is reduced in the ``noise dominated" regime ($\Sigma<\Sigma_i$, upper data points in Fig.~\ref{fig_fluctuations_sigma}(c)). This asymmetry is likely to be due to the different origins of mechanical noise in these two regimes, where it arises mainly due to activated events in the ``noise dominated" regime.

\begin{figure*}[t]
\centering
\includegraphics[width=1.9\columnwidth, clip]{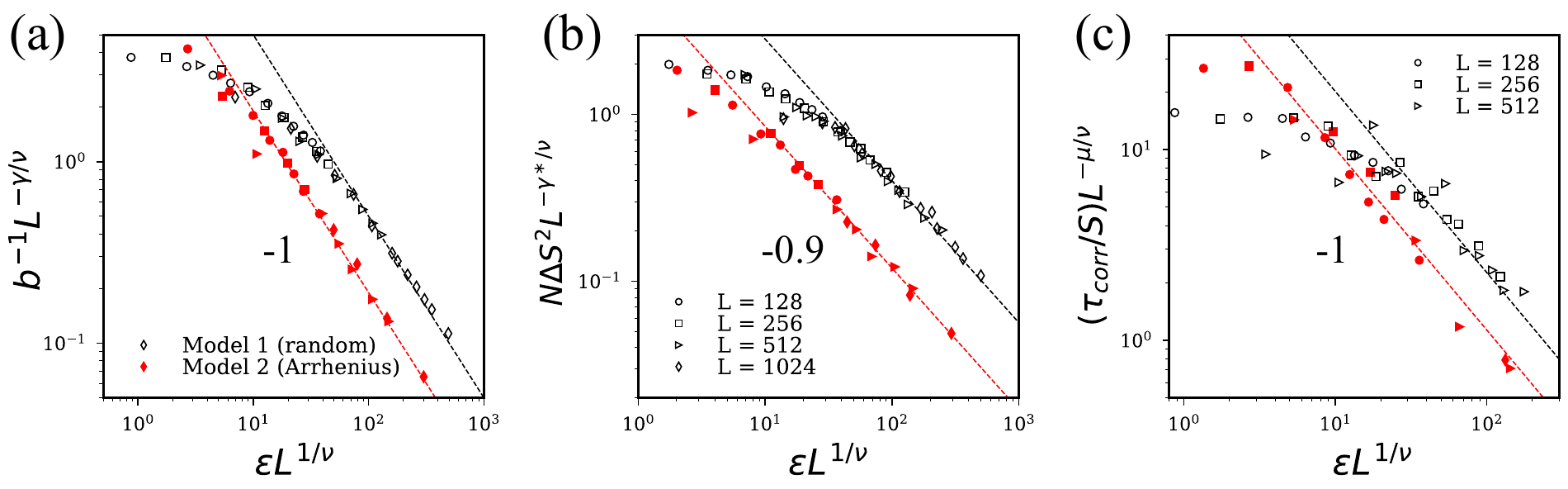}
\caption{(a) Susceptibility $\chi=1/b$ rescaled by $L^{-\gamma/\nu}$ as a function of the scaled distance to the critical point $\varepsilon L^{1/\nu}$ for model 1 (unfilled black symbols) and model 2 (solid red symbols). (b) Rescaled variance $\Delta S^2$ vs.~$\varepsilon L^{1/\nu}$. (c) Rescaled correlation time vs.~$\varepsilon L^{1/\nu}$}
\label{fig_FFS_fluctuations_2models}
\end{figure*}

\paragraph*{Varying the noise magnitude $\varepsilon$:}

We now investigate the scaling of fluctuations at the inflexion point of the flow curve ($\Sigma = \Sigma_i$) when varying the noise magnitude $\varepsilon$. We compute the variance of the fluctuations and extract their correlation time from an exponential fit of the auto-correlation of $S$. 
A finite size data collapse is performed using $\nu =1$. 
We find, for the two models of noise, a power law increase of both the variance (Fig.~\ref{fig_FFS_fluctuations_2models}(b)) and the correlation time (Fig.~\ref{fig_FFS_fluctuations_2models}(c)) of the fluctuations when approaching the critical point, with the exponents $\gamma^* \simeq 0.9$ and $\mu \simeq 1$ respectively.

Let us now discuss the scaling of shear rate distributions depicted in Fig.~\ref{Fig_giant_fluctuations}(d) in the light of the critical exponents estimated in this section.
In Fig.~\ref{Fig_giant_fluctuations}(d), the shear rate distributions at the transition could be approximately collapsed by rescaling the shear rate with a factor $N^{x}$ ($N=L^2$) and the best collapse was found with an exponent $x=0.275$.
In other words, the width of the distribution of shear rate scales as: $\Delta \dot\gamma \sim L^{-2x} \sim L^{-0.55}$.
For sufficiently small fluctuations of the shear rate, the fluctuations of $S=\log(\dot\gamma)$, $\Delta S$, correspond approximately to relative fluctuations of $\dot\gamma$, $\Delta \dot\gamma/\dot\gamma_c$, with $\dot\gamma_c$ the critical shear rate (approximately independent of system size, see Appendix C). 
Hence similar scalings for $\Delta S$ and $\Delta \dot\gamma$ with the system size would be expected.
From the data collapse of Fig.~\ref{Fig_giant_fluctuations}(d), the scaling for the shear rate variance is $L^2\Delta \dot\gamma^2 \sim L^{0.9}$. This is consistent with the scaling of Fig.~\ref{fig_FFS_fluctuations_2models}(b), where data collapse for the variance of $S$ suggests a scaling form: $L^2\Delta S ^2 \sim L^{\gamma^*/\nu}$, with $\gamma^*/\nu=0.9$.

Note that, due to finite time limitations of our simulations in the critical region, the data for the correlation time of the fluctuations (Fig.~\ref{fig_FFS_fluctuations_2models}(c)) is restricted to intermediate system sizes ($L\leq 512$) and exhibit strong scattering (see Appendix B, Fig.~\ref{fig:Appendix_corr}(a)). An approximate collapse can still be performed with an exponent $\mu = 1.0 \pm 0.2$ and seems to be independent of the model of noise. 
As noted in \cite{legoff2019critical}, this value would correspond to a dynamic scaling exponent $z=\mu/\nu \approx 1$, far from the equilibrium mean-field value $z=2$ obtained for non-conserved scalar order parameters \cite{HohenbergRMP1977}.

\section{Conclusion}

In this work, we studied two models of a fluidising noise leading to a transition between a self-fluidised regime and an externally fluidised regime in the flow of soft glassy materials.
Upon increase of the external noise amplitude we evidence the vanishing of shear bands, which is an indication of a transition from an unstable shear-localised flow to a stable homogeneous flow. We find that a decrease of the fluidising noise amplitude in the homogeneous flow regime yields increasingly large and long lived fluctuations of the macroscopically measurable shear rate. In this scenario a single trajectory of the system can strongly differ from its average behaviour as given by the constitutive flow curve, thus leading to several difficulties for experimental characterisations of the flow behaviour. 

Since the correlation time of the fluctuations becomes increasingly large at the transition point, averaging values in the steady state becomes tedious and care has to be taken in the data interpretation. This is a situation where conventional continuum descriptions of the flow tend to break down and call for more sophisticated modelling approaches that allow for incorporating the spatio-temporal features of the fluctuations in the rheologically relevant quantities.

Another important message regarding experimental and numerical studies is that, when using shear-imposed protocols, the flow can be heterogeneous due to the non-monotonicity of the constitutive flow curve. In the vicinity of the transition, the time for the shear bands to coarsen becomes increasingly large, and the steady state of the system may be out of reach on any experimentally relevant timescale. 

All of the above features can be accounted for in the framework of non-equilibrium phase transitions, as resulting from an out-of-equilibrium critical transition point.
The properties of this critical point, studied in detail for the random activation model in a previous work \cite{legoff2019critical} appear to remain unchanged for the Arrhenius activation model. 
While the rheology at low stress differs between the two models, we find a generic scaling of the flow curves in the critical regime that is well fitted by a Landau-type expansion, as in \cite{wortel2016criticality}. 
The scaling of the order parameter (power law scaling with an exponent $\beta \simeq 0.5$), as determined from the flow profiles in the phase coexistence regime, is consistent with the scaling of the flow curves in the stable flow regime.
We also find power law scalings of the susceptibility and the shear rate fluctuations, with exponents being independent of the model of noise.
In conclusion, the finite shear-rate critical point studied here in a minimal elasto-plastic model suggests that a generic critical behaviour arises in systems combining a non-monotonic flow curve with a fluidisation process, irrespective of the detailed physical mechanisms at play. 

We discussed the consequences of competing self-fluidisation and external fluidisation mechanisms on the rheology of soft glassy materials and in the following we would like to suggest situations where this phenomenon might be of importance.
As shown experimentally, this transition can easily arise in frictional granular materials \cite{wortel2016criticality}. But one could also speculate about other systems where such mechanisms could be at play. 
The minimal ingredients for the emergence of this critical point in systems sheared at a finite strain rate are: (i) a microscopic mechanism at the origin of self-fluidisation (such as an intrinsic timescale for restructuration in the material \cite{coussot2010physical}), and (ii) a source of mechanical noise independent of the flow.
An example for a class of materials, that provides naturally various sources of additional mechanical noise with a fluidisation effect, are dense active systems, such as biological tissues \cite{matoz2017nonlinear,mandal2016active,tjhung2017discontinuous}. Depending on the details of microscopic interactions, self-fluidisation could arise from a competition between intrinsic timescales in the system and shear; hence such systems would be good candidates for the emergence of critical dynamics accompanied by giant fluctuations as described in this work.

\subsection*{Acknowledgments}
KM and MLG acknowledge funding from the Centre Franco-Indien pour la Promotion de la Recherche Avanc\'ee  (CEFIPRA) Grant No.\ 5604-1 (AMORPHOUS-MULTISCALE). KM acknowledges financial support of the French Agence Nationale de la Recherche (ANR), under grant ANR-14-CE32-0005 (FAPRES). Most of the computations were performed using the Froggy platform of the CIMENT infrastructure supported by the Rh\^one-Alpes region (GRANTCPER07-13 CIRA).

\appendix{}

\begin{figure*}
\centering
\includegraphics[width=1.4\columnwidth, clip]{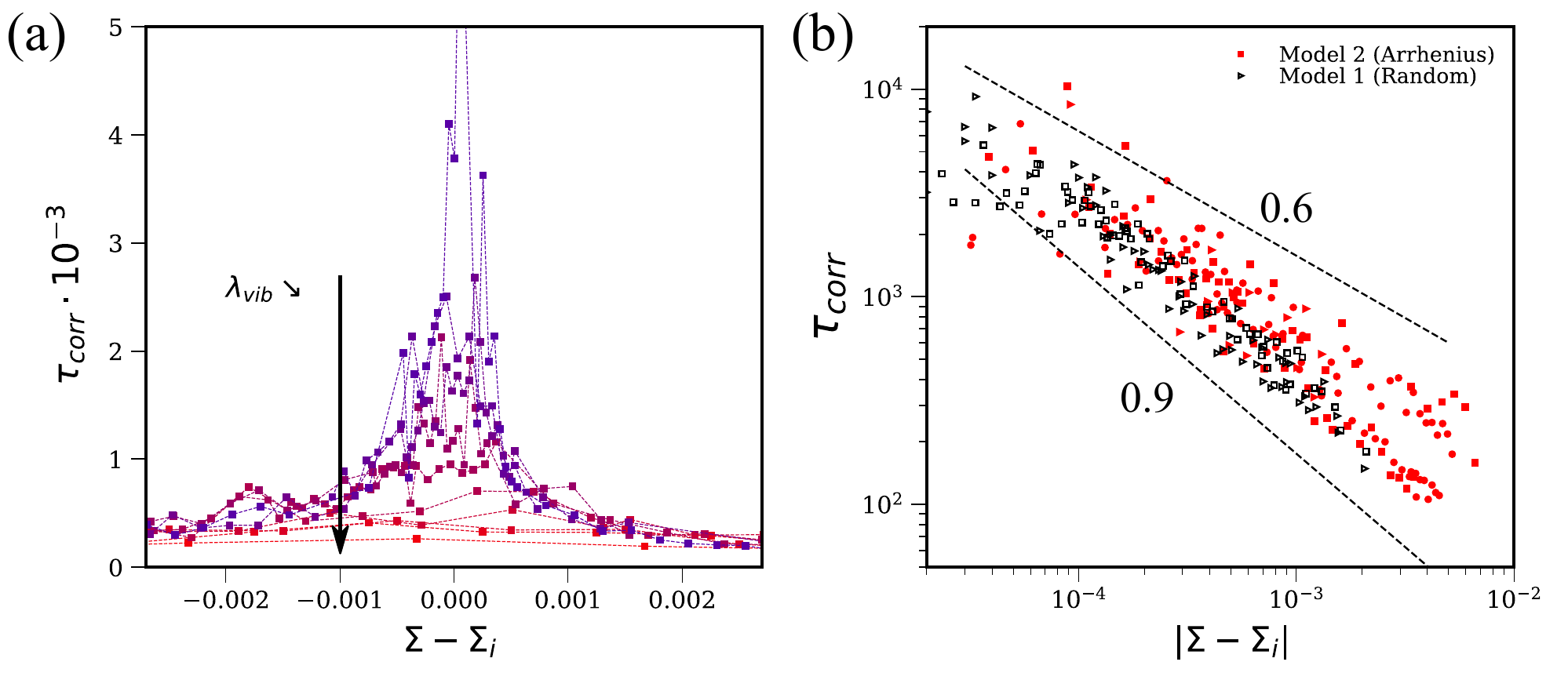}
\caption{{(a) Auto-correlation timescale $\tau_{corr}$} as a function of $\Sigma-\Sigma_i$ for various values of noise magnitude $\lambda_{vib}$ (model 2). (b) Log-log plot of the correlation time $\tau_{corr}$ as a function of $|\Sigma-\Sigma_c|$ for $\varepsilon \simeq 0$ ("critical isotherm") for the two models of noise.}
\label{fig:Appendix_corr}
\end{figure*}

\section{Data analysis}
We perform simulations of the 2d elasto-plastic model using shear- and stress-controlled protocols.

Using a shear-imposed protocol, we measure the average steady-state stress in the system to compute the flow curve, by averaging over a strain window $\gamma = 50$. To compute the shear rate profiles in the shear banding regime, we average the profiles along the direction in which the flow is homogeneous and over a strain window $\gamma = 1000$.

Using a stress-controlled protocol, we analyse time-series of the average shear rate in the system (of average duration $T=2.10^6$ for $L=128$ and $L=256$, $T=6.10^5$ for $L=512$ and $T=10^5$ for $L=1024$, corresponding to strains ranging from $\gamma=2000$ to $4.10^4$). We compute  $S = \log(\dot\gamma)$, as well as its variance $\Delta S^2$.

\section{Temporal autocorrelation}

To extract the correlation time $\tau_{corr}$, we fit the auto-correlation function 
$C(\tau)=\langle \Delta S(t+\tau)\Delta S(t)\rangle$ to an exponential and extract the characteristic time $\tau_{corr}$ from the fit.
We report in Fig.~\ref{fig:Appendix_corr}(a) the correlation time as a function of the distance (in stress) to the inflexion point of the flow curve for various values of $\lambda_{vib}$ (model 2), which exhibits a sharp peak at the critical point. 
We report in a log-log plot the correlation time as a function of the absolute distance to the inflexion point $|\Sigma-\Sigma_i|$, for a noise magnitude close to the critical point ($\varepsilon \simeq 0$), for the two models of noise. 
Although the data is more noisy than for the variance of the fluctuations (see main text, Fig.~\ref{Fig6_fluctuations_stress}), we can see there is a power-law scaling of the correlation time as well, with an exponent close to that of the variance.

\section{Landau expansion fit}

\begin{figure}[b]
\centering
\includegraphics[width=0.75\columnwidth, clip]{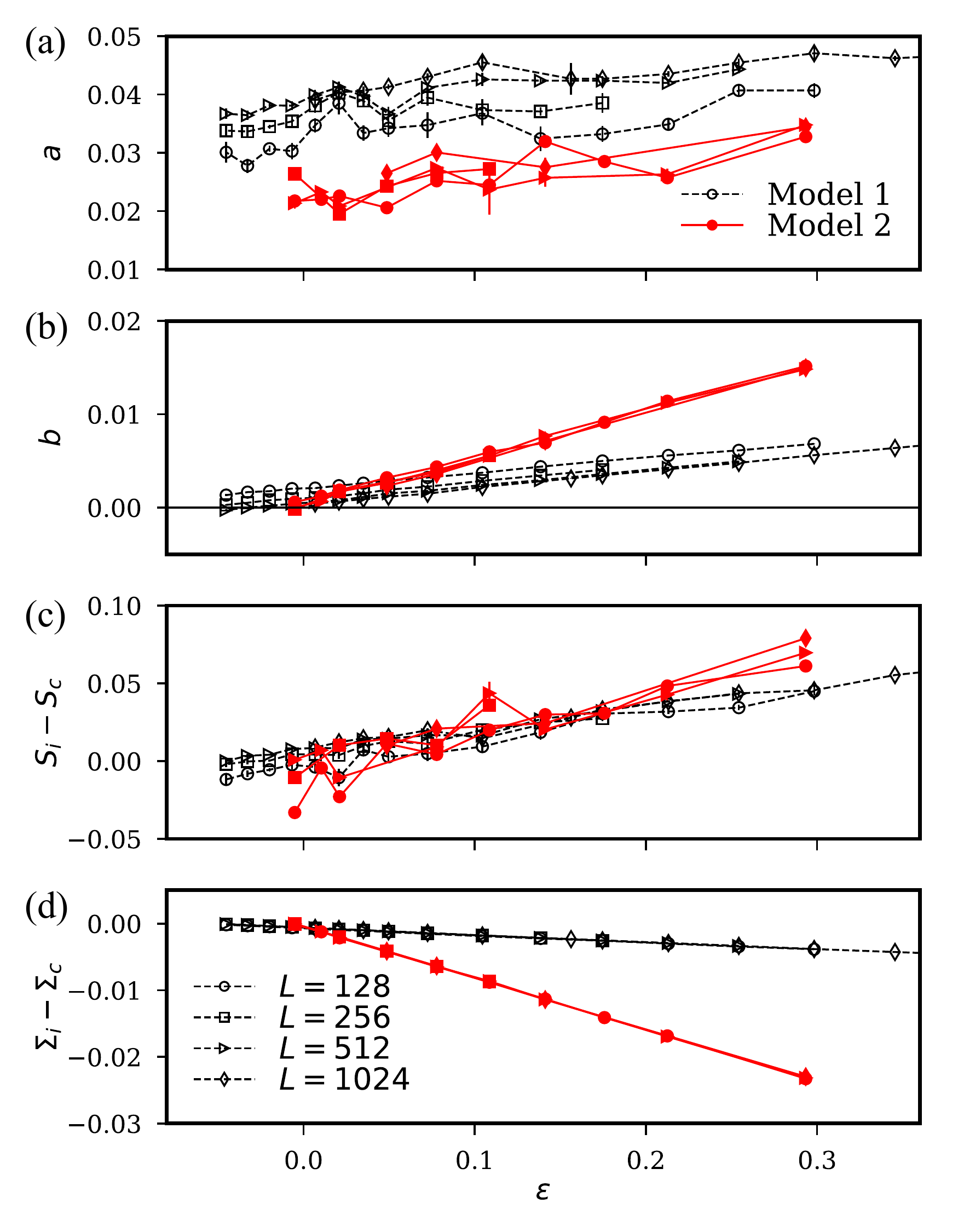}
\caption{Parameters from the Landau-like expansion fit for various values of the noise $\varepsilon$ and various system sizes $L$, for model 1 (black empty squares, dashed line) and model 2 (red dot, solid line). (a) Prefactor $a$; (b) inverse susceptibility $b$; (c) and (d) coordinates of the inflexion point shifted by the critical point location, respectively $S_i-S_c$ (c) and $\Sigma_i - \Sigma_c$ (d).}
\label{fig_fitting_params}
\end{figure}

Fig.~\ref{fig_fitting_params} depicts the fitting parameters from the Landau-like expansion fit (Eq.~\ref{fit_Landau}): $a(\varepsilon)$, $b(\varepsilon)$, $S_i(\varepsilon)$ and $\Sigma_i(\varepsilon)$ for the two models of noise.


%

\end{document}